\documentclass[runningheads]{llncs}
\usepackage{graphicx}
\usepackage{subcaption}
\usepackage{float}

\usepackage{url}
\begin{document}
\title{Enhancement of Dysarthric Speech Reconstruction by Contrastive Learning}
%
%
\author{Fatemeh Keshvari \and
	Rahil Mahdian Toroghi \and
	Hassan Zareian
}
\institute{Iran Broadcasting University (IRIBU), Tehran, Iran 	
	\\
	\email{\{fkeshvari4955@gmail.com\}},	\email{\{mahdiantr1974@ieee.org\}},
	\email{\{zareian@iribu.ac.ir\}}
}
\maketitle              
\begin{abstract}
Dysarthric speech reconstruction is challenging due to its pathological sound patterns. Preserving speaker identity, especially without access to normal speech, is a key challenge. Our proposed approach uses contrastive learning to extract speaker embedding for reconstruction, while employing XLS-R representations instead of filter banks. The results show improved speech quality, naturalness, intelligibility, speaker identity preservation, and gender consistency for female speakers. Reconstructed speech exhibits 1.51 and 2.12 MOS score improvements and reduces word error rates by 25.45$\%$ and 32.1$\%$ for moderate and moderate-severe dysarthria speakers using Jasper speech recognition system, respectively. This approach offers promising advancements in dysarthric speech reconstruction.

\end{abstract}
\section{Introduction}
Dysarthria, a speech disorder arising from muscular weakness affecting speech articulators such as the lips, tongue, and larynx, manifests as a distortion of speech characterized by irregular acoustic patterns \cite{yunusova2008articulatory}. In the endeavor to reconstruct dysarthric speech, the preservation of the speaker's unique identity poses a formidable challenge, primarily stemming from the practical complexities inherent in obtaining a patient's unaffected speech. Consequently, the imperative of our research lies in the development of methodologies that facilitate the reconstruction of dysarthric speech while simultaneously safeguarding the distinct vocal identity of dysarthric individuals, thereby contributing to an improved quality of life.

In this work, we present a novel approach for dysarthric speech reconstruction leveraging cascaded automatic speech recognition (ASR) and text-to-speech (TTS) models. Our method involves initial transcription of the input speech through an ASR model, followed by the utilization of the transcribed linguistic content and a speaker-specific embedding vector by the TTS model to generate the reconstructed speech. To address the challenge of extracting speaker embedding vectors from dysarthritic speech using a pre-trained speaker encoder, we introduce a novel application of contrastive learning.

Moreover, training the ASR model is complicated by the inherent difficulty of dysarthric speakers in pronouncing words accurately, leading to increased reconstruction errors. To mitigate this issue, we employ cross-lingual self-supervised speech representation wav2vec 2.0 as an input feature for ASR model training, inspired by \cite{hernandez2022cross}.

In the context of dysarthric speech reconstruction, statistical methods such as non-negative matrix factorization (NMF) \cite{aihara2012consonant}, and partial least squares  \cite{aihara2017phoneme}, or deep learning methods such as sequence-to-sequence models  \cite{doshi2021extending}, gated convolution networks \cite{chen2020enhancing} and knowledge distillation \cite{wang2020end} exhibit limitations in  preserving speaker identity. Several strategies have emerged to address this problem in dysarthric speech reconstruction including the utilization of a controller network in  \cite{chen2018generative}, that serves to extract both linguistic content and speaker-specific characteristics from dysarthric speech. This network is trained using perceptual similarity and high-level feature comparison.

\section{Related Works}
NED\footnote{Neural Encoder-Decoder}-based architectures have exhibited significant improvements in the intelligibility of reconstructed speech when compared to their generative adversarial network (GAN) counterparts. However, they suffer from training challenges and inefficiencies caused by the cascaded pipeline of the content encoder.

Conversely, GAN-based DSR\footnote{Dysarthric speech recognition} systems implicitly model the linguistic and prosodic aspects of speech. This implicit modeling approach, while potentially advantageous in certain contexts, renders the identification and analysis of reconstruction errors more complex and less tractable.

To preserve speaker identity in dysarthric speech reconstruction, a two-stage approach is proposed in  \cite{huang2021preliminary}. Initially, the dysarthric speech input is converted into reference speaker's normal speech through a sequence-by-sequence model. Subsequently, the patient's identity is recovered using a VAE\footnote{Variational Auto-Encoder}-based non-parallel voice conversion model. As described in  \cite{chen2022phoneme}, a speaker encoder is employed to extract speaker traits, alongside a text encoder, and joint training of text and speech to improve the intelligibility of  dysarthric speech.

In the speaker verification based dysarthric speech reconstruction SV-DSR \cite{wang2020learning}, a pre-trained speaker encoder enhances dysarthric speech intelligibility with prosody correction, although it struggles to accurately capture speaker features. This particularly happens for female patients, so that they are prone to voice gender alterations during the reconstruction process.

To address this issue, an innovative approach known as Adversarial Speaker Adaptation (ASA-DSR) is introduced \cite{wang2022speaker}. ASA-DSR leverages an adaptation loss function to fine-tune the speaker encoder, enabling it to effectively extract speaker embedding vectors from dysarthric speech. Furthermore, adversarial learning is employed to maximize the similarity between the outputs of these models, thereby preserving the effectiveness of the SV-DSR model while mitigating gender-related voice variations in reconstructed speech.

Liu et al. \cite{liutwo2024}, have introduced a novel approach by proposing a two-stage methodology using voice conversion method. In the first stage a same-gender-retrieval strategy has been leveraged to match the dysarthric speech with the normal speech of the same gender. In the second stage, they try to improve the quality of the  repaired speech.

While preserving speaker identity, these methods yield promising results in dysarthric speech reconstruction, yet they still exhibit substantial disparities from normal speech in terms of intelligibility and naturalness.

Recently, Wang et al. \cite{wang2024unit}, proposed a Unit-DSR system that leverages the HuBERT domain-adaptation capability to constrain the speech content restoration. It avoids the cascaded sub-modules and achieves a remarkable results in terms of WER\footnote{Word Error Rate}.

\begin{figure}[!t]
	\centering
	\fbox{\includegraphics[width=1\linewidth, height=.65\textheight]{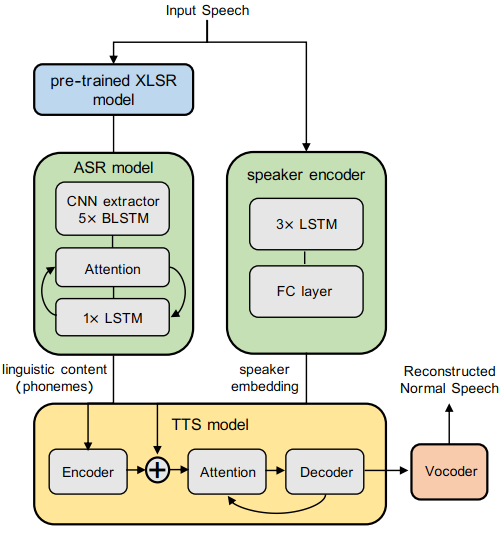}}
	\caption{Functional Diagram of the proposed model}
	\label{fig:fig1proposeddiagram}
	\vspace{-2mm}
\end{figure}

\section{The Proposed Model}

The proposed model, depicted in Figure \ref{fig:fig1proposeddiagram}, comprises three key components: a speaker encoder, an ASR model, and a TTS model. The speaker encoder derives a speaker's embedding vector, encapsulating their distinctive characteristics for preserving speaker identity during dysarthric speech reconstruction. Simultaneously, the ASR model extracts linguistic content from the speech. Subsequently, the TTS model utilizes both the linguistic content and the extracted speaker embedding vector from the dysarthric speech, resulting in speech output that faithfully retains the patient speaker's identity while enhancing naturalness and intelligibility.

\subsection{Speaker Encoder}

We employ a pre-trained speaker encoder network trained on standard speech data using generalized end-to-end loss \cite{wan2018generalized}. However, due to substantial distinctions between normal and dysarthric speech, this encoder may not consistently produce appropriate embedding vectors, particularly in cases involving female speakers.

Dysarthria influences the patient's speech production system, leading to a reduction in vocal cord vibration frequency and, consequently, a lower pitch in female voices  \cite{rudzicz2011production}. This vocal change caused by the disease means that the pre-trained speaker encoder may alter voice gender and fail to preserve the speaker's identity in the reconstructed speech.

To address this challenge, our proposed solution involves simulating pitch shifting using normal speech and employs contrastive learning via a triple loss function to refine the parameters of the speaker encoder. The triple loss function divides the training data into three categories: an anchor, a positive, and a negative sample. The anchor sample serves as a reference for learning its embedding vector, while the positive sample corresponds to a similar example, and the negative one represents a dissimilar one. This triple loss function, guides the training process in achieving the desired outcomes, as in  \cite{chung2020defence}:\vspace{-4mm}

\begin{equation}
L(\tau)\! =\! \sum \mathbf{max}\big(\bigg|\bigg|f(x_a^{\tau})-f(x_p^{\tau}) \bigg|\bigg|_2^2\!-\!\bigg|\bigg|f(x_a^{\tau})-f(x_n^{\tau}) \bigg|\bigg|_2^2\!+\!\alpha,0\big)
\label{eq:eq1}
\end{equation}
where $\tau$ represents the desired triplet, $f(x_a^{\tau})$ denotes the embedding of the anchor, and $f(x_p^{\tau}), f(x_n^{\tau})$ denote the embedding vectors of positive and negative samples, respectively.

In order to address the issue of voice gender alteration in the reconstructed speech, we employ pitch shifting to simulate a more bass-like voice for the patient speaker, serving as a negative sample in the triple loss function. Additionally, due to the extended duration of dysarthric speech compared to normal speech, tempo-changing method is applied to positive samples.

Given potential ineffectiveness for simulating the negative sample with male patients, we exclusively use tempo-changing to simulate the positive sample for fine-tuning the speaker encoder parameters.

\subsection{The ASR Model}
Our ASR model architecture, based on \cite{hori2017advances}, combines attention and connectionist temporal classification (CTC) for enhanced speech recognition. Due to limited dysarthric speech data, we adopt a two-stage training process: 1) initial network pre-training with normal speech data, and 2) subsequent fine-tuning of parameters using speaker-specific dysarthric speech data.

 The ASR model is optimized by minimizing a weighted sum of sequence-by-sequence and CTC loss functions \cite{hori2017advances,graves2006connectionist}.
\begin{equation}
L_{ASR} =\! \lambda_{s2s}L_{s2s}+(1-\lambda_{s2s})L_{ctc}
\end{equation}
where $L_{s2s}$ denotes the cross-entropy between the decoder outputs from the ASR model and the ground-truth character label sequence, and $L_{ctc}$ denotes the cross-entropy between all possible sequences mapped to the ground-truth character label sequence by inserting blanks and repeating characters. $\lambda_{s2s}$ controls the weights of the model for multi-task learning between sequence-by-sequence and CTC approaches and is empirically set to 0.5. 

The ASR model initially used Mel-Spectrogram input in its structure. However, in this study, we enhance dysarthric speech recognition and subsequently reduce speech reconstruction errors by employing the XLS-R model \cite{conneau2020unsupervised}, based on the wav2vec2.0 architecture trained on multilingual data \cite{baevski2020wav2vec}, as a feature extractor.

Evaluations indicate improved performance, particularly in recognizing dysarthric speech, attributed to its inclusion of a diverse set of similar phonemes.

\subsection{The TTS Model}
In text-to-speech synthesis, recent advancements employ attention-based seq-to-seq\footnote{Sequence-to-Sequence} models to generate high-quality speech. This study leverages the Tacotron model \cite{wang2017tacotron}, for text-to-speech conversion. Given that the dataset comprises solely isolated words articulated by individuals with dysarthria, Tacotron emerges as a judicious and adequate choice for text-to-speech (TTS) synthesis. While alternative more recent TTS architectures demonstrate comparable efficacy, Tacotron's capabilities align well with the specific constraints and characteristics of the dataset in question.

 Given the unavailability of typical speech data from individuals affected by dysarthria, a zero-shot scenario compels us to adopt a speech generation approach that relies on speaker embedding vectors extracted by a speaker encoder.


To address this challenge, a transfer learning method originating from speaker verification tasks, as in \cite{jia2018transfer}, has been adopted. This approach involves augmenting the encoder output of the text-to-speech model at each time step with the target speaker's embedding vector. By concatenating these vectors and utilizing them as inputs for the attention layer, the model converges effectively in multi-speaker scenarios \cite{jia2018transfer}. Consequently, Tacotron predicts speech mel-spectrograms based on input phonemes and the speaker's embedding vector, while the WaveRNN vocoder \cite{kalchbrenner2018efficient},  converts the spectrogram into speech waveforms in the time domain.

\section{Experiments}

\subsection{Experimental Setting}

The speaker encoder was pre-trained using the VoxCeleb1 and VoxCeleb2 datasets \cite{nagrani2017voxceleb,chung2018voxceleb2}, followed by fine-tuning with  LibriSpeech (train-clean-100) dataset \cite{panayotov2015librispeech}.

 For training the TTS model and vocoder, the VCTK dataset was employed \cite{veaux2016superseded}. The ASR model, on the other hand, was trained using the UASpeech dataset \cite{kim2008dysarthric}, encompassing 13 healthy speakers and 15 speakers with cerebral palsy. Each speaker contributed three speech blocks, each containing 225 utterance files encompassing numbers, computer commands, common, and uncommon words.

Given the limited availability of speech data for each patient, the ASR model was pre-trained using the speech data from healthy speakers and subsequently fine-tuned with dysarthric speaker data. For training and testing, we utilized speech data from blocks 1 and 3 of each speaker, reserving block 2 for evaluation. This study involved two female speakers (F02, F04) and two male speakers (M05, M07), each presenting with moderate to moderate-severe dysarthria.

The speaker encoder comprises a 3-layer LSTM with 256 dimensions and a fully connected layer for extracting embedding vectors. Pre-training follows a method akin to that detailed in \cite{wan2018generalized}. In cases of moderate-severe dysarthria, pitch shifting and tempo-changing coefficients are set to 0.5, while for moderate dysarthria, these coefficients are adjusted to 0.25 and 0.7, respectively, during the fine-tuning process for simulating positive and negative samples in the triple loss function. This fine-tuning of speaker encoder parameters involves 5,000 iterations with a batch size of 64. 

The ASR model, illustrated in Figure \ref{fig:fig1proposeddiagram}, comprises an encoder consisting of a convolutional extractor and a 5-layer bidirectional LSTM (BLSTM), a location-aware attention module  \cite{chorowski2015attention}, and a decoder with a single-layer LSTM. During pre-training and fine-tuning, the model is trained with batch sizes of 16 for 100,000 and 10,000 iterations, respectively. To account for increased speech duration caused by dysarthria, data augmentation was implemented during pre-training. The model utilizes XLS-R speech representation as input, where the XLS-R model is trained on speech data from 53 languages, totaling 56,000 hours of speech.

This study, inspired by  \cite{jia2018transfer}, employs transcribed text data and corresponding speech for training a Text-to-Speech (TTS) model. It focuses on expediting convergence and refining the pronunciation of uncommon words. The process involves normalizing the text data and encoding phonemes into 256-dimensional embedding vectors. The model's training comprises 600K iterations with a batch size of 12.

\subsection{Experimental Results and Analytics}
This work conducts subjective and objective assessments to gauge the efficiency of the proposed dysarthric speech reconstruction method in enhancing speech quality. Baseline models SV-DSR and ASA-DSR are employed for comparison. Some sample results are available online\footnote{\url{<https://fatemehkshvr.github.io/>}}.

\subsubsection{Subjective Evaluation}

We employed a 5-point Mean Opinion Score (MOS) test to assess our dysarthric speech reconstruction method. This evaluation involved 20 listeners, encompassing both speech processing experts and non-experts. Each listener reviewed 20 randomly chosen reconstructed speech files, alongside the original dysarthric speech, rating them on a scale from 1 (bad) to 5 (excellent) based on two criteria: speech naturalness and speaker similarity.

Speech reconstruction seeks to enhance input speech to closely resemble natural speech. In the speech naturalness MOS test, listeners initially evaluate dysarthric speech with scores from 1 to 5. Subsequently, they assess the corresponding reconstructed speech using the same scoring method. Table \ref{tbl:t1} presents the test results.

\begin{table}[h]
	\caption{MOS results for Speech Naturalness with 95$\%$ confidence intervals.}\vspace{2mm}
	\label{tbl:t1}
	\centering
	\resizebox{0.9\columnwidth}{!}{%
		\begin{tabular}{l|l|l|l|l}
			\textbf{Speech}        & \textbf{F04}           & \textbf{M05}           & \textbf{F02}           & \textbf{M07}           \\ \hline
			\textbf{Original}      & 2.86 $\pm$ 0.08 & 2.53 $\pm$ 0.09 & 1.84 $\pm$ 0.14 & 2.08 $\pm$ 0.09 \\ \hline
			\textbf{Reconstructed} & \textbf{4.15} $\pm$ 0.09 & \textbf{4.08} $\pm$ 0.05 & \textbf{4.21} $\pm$ 0.07 & \textbf{3.95} $\pm$ 0.12
		\end{tabular}%
	}
\end{table}

The proposed approach demonstrates successful performance in dysarthric speech reconstruction, particularly in enhancing speech naturalness, as indicated by improvements in MOS test scores. Notably, when the patient speaker is female, the reconstructed speech exhibits higher quality. This improvement can be attributed to pitch shifting for simulating female voices and the application of contrastive learning. It's noteworthy that in the proposed approach, data augmentation via tempo adjustment is used for reconstructing dysarthric speech with male speakers.

\noindent
The objective of this study is to maintain speaker identity during speech reconstruction, as previously outlined. Therefore, it is crucial to assess the similarity between the reconstructed and original dysarthric speech concerning speaker identity. To achieve this, MOS tests and human listeners were employed, with Table \ref{tbl:t2} presenting the outcomes. The findings reveal that the contrastive learning approach effectively preserves speaker identity, particularly in addressing voice gender changes in reconstructed speech when the speaker is female. Notably, evaluating this aspect can be challenging due to the absence of normal patient speech, potentially influencing MOS scores despite the substantial dysarthric-normal speech distinction.
\begin{table}[h]
	\centering
	\caption{MOS results for Speaker Similarity with 95$\%$ confidence intervals.}\vspace{2mm}
	\label{tbl:t2}
	\resizebox{0.9\textwidth}{!}{%
		\begin{tabular}{lllll}
			\multicolumn{1}{l|}{\textbf{Approaches}} & \multicolumn{1}{l|}{\textbf{F04}}           & \multicolumn{1}{l|}{\textbf{M05}}           & \multicolumn{1}{l|}{\textbf{F02}}           & \textbf{M07}           \\ \hline
			\multicolumn{1}{l|}{\textbf{SV-DSR}}     & \multicolumn{1}{l|}{2.27 $\pm$ 0.10} & \multicolumn{1}{l|}{2.70 $\pm$ 0.08} & \multicolumn{1}{l|}{1.88 $\pm$ 0.13} & 2.55 $\pm$ 0.14 \\ \hline
			\multicolumn{1}{l|}{\textbf{ASA-DSR}}    & \multicolumn{1}{l|}{3.16 $\pm$ 0.15} & \multicolumn{1}{l|}{3.27 $\pm$ 0.10} & \multicolumn{1}{l|}{2.93 $\pm$ 0.15} & 3.20 $\pm$ 0.13 \\ \hline
			\multicolumn{1}{l|}{\textbf{Proposed}} &
			\multicolumn{1}{l|}{\textbf{3.55} $\pm$ 0.16}  &
			\multicolumn{1}{l|}{\textbf{3.48} $\pm$ 0.09}  &
			\multicolumn{1}{l|}{\textbf{3.32} $\pm$ 0.08}  & \textbf{3.28} $\pm$ 0.13
		\end{tabular}%
	}
\end{table}

\subsubsection{Objective Evaluations}

Dysarthria significantly impairs speech intelligibility, particularly in terms of word pronunciation. To assess the effectiveness of our proposed approach in enhancing dysarthric speech intelligibility, we employed the word error rate (WER) in conjunction with the Jasper speech recognition system \cite{li2019jasper}. The outcomes of this evaluation are presented in Table \ref{tbl:t3}. Comparing the results, we can infer that training ASR models with XLS-R speech representation, rather than filterbanks, yields improved accuracy in dysarthric speech recognition and reconstruction.


\begin{table}
	\centering
	\caption{WER ($\%$) results for Speech Intelligibility}\vspace{2mm}
	\label{tbl:t3}
	\resizebox{0.55\textwidth}{!}{
	\begin{tabular}{|l|l|l|l|l|} 
		\hline
		\textbf{Approaches}             & \textbf{F04 }                 & \textbf{M05}                  & \textbf{F02}                  & \textbf{M07}                   \\ 
		\hline
		\textbf{Original}             & 81                   & 91                   & 96                   & 95                    \\ 
		\hline
		\textbf{SV-DSR}               & 64.6                 & 61.7                 & 65.3                 & 62.7                  \\ 
		\hline
		\textbf{ASA-DSR}              & 65.6                 & 62.5                 & 65.8                 & 62.7                  \\ 
		\hline
		\textbf{Proposed}             & \textbf{61.9}                 & \textbf{59.2}                 & \textbf{63.7}                 & \textbf{63.1}                  \\ 
		\hline
		\multicolumn{1}{l}{} & \multicolumn{1}{l}{} & \multicolumn{1}{l}{} & \multicolumn{1}{l}{} & \multicolumn{1}{l}{} 
	\end{tabular}
}
\end{table}

Subjective and objective evaluations indicate that the proposed approach effectively preserves the speaker's identity while enhancing the naturalness and intelligibility of reconstructed speech. Compared to previous methods, our model produces higher-quality speech output in terms of speaker similarity. Although the Mean Opinion Score (MOS) results demonstrate significant improvements in the naturalness of reconstructed speech over dysarthric speech, it still falls short of matching normal speech quality. Similarly, while our model achieves better performance with a lower Word Error Rate (WER), the WER remains high, partly due to evaluation errors in Jasper's assessment. This highlights the importance of subjective evaluations, which can provide more accurate and reliable results, particularly in assessing speech intelligibility. Therefore, listening tests may be necessary for a comprehensive evaluation of intelligibility. Nonetheless, objective evaluations, based on standardized criteria, will remain crucial in assessing the effectiveness of the proposed method.

\section{Conclusion}
This study presents and evaluates a novel methodology leveraging contrastive learning for the optimization of speaker encoders in the context of dysarthric speech reconstruction. The proposed approach effectively mitigates voice gender alterations, such as the pathological lowering of pitch in female speakers, by incorporating pitch-shifted samples as negative exemplars within a triplet loss framework. Concurrently, the implementation of data augmentation techniques, specifically tempo adjustment during model training, significantly improves the perceptual fidelity of the reconstructed speech.
Furthermore, this investigation assesses the efficacy of automatic speech recognition (ASR) models trained on XLS-R representations. The findings demonstrate that utilizing these representations as input features substantially reduces reconstruction errors in dysarthric speech and yields lower word error rates in subsequent evaluations.
The methodology introduced herein represents a significant advancement in the field of speech pathology and rehabilitation, offering potential applications in both diagnostic and therapeutic contexts. By addressing the challenges associated with gender-specific voice alterations and improving overall speech reconstruction quality, this approach paves the way for more accurate and personalized interventions in cases of dysarthria and related speech disorders.

%
%
%
%
%
 \bibliographystyle{splncs04}
 \bibliography{Medprai2024-Keshvari}

\begin{thebibliography}{10}
\providecommand{\url}[1]{\texttt{#1}}
\providecommand{\urlprefix}{URL }
\providecommand{\doi}[1]{https://doi.org/#1}

\bibitem{aihara2012consonant}
Aihara, R., Takashima, R., Takiguchi, T., Ariki, Y.: Consonant enhancement for
  articulation disorders based on non-negative matrix factorization. In:
  Proceedings of The 2012 Asia Pacific Signal and Information Processing
  Association Annual Summit and Conference. pp.~1--4. IEEE (2012)

\bibitem{aihara2017phoneme}
Aihara, R., Takiguchi, T., Ariki, Y.: Phoneme-discriminative features for
  dysarthric speech conversion. In: Interspeech. pp. 3374--3378 (2017)

\bibitem{baevski2020wav2vec}
Baevski, A., Zhou, Y., Mohamed, A., Auli, M.: wav2vec 2.0: A framework for
  self-supervised learning of speech representations. Advances in neural
  information processing systems  \textbf{33},  12449--12460 (2020)

\bibitem{chen2020enhancing}
Chen, C.Y., Zheng, W.Z., Wang, S.S., Tsao, Y., Li, P.C., Lai, Y.H.: Enhancing
  intelligibility of dysarthric speech using gated convolutional-based voice
  conversion system. In: Interspeech. pp. 4686--4690 (2020)

\bibitem{chen2018generative}
Chen, L.W., Lee, H.Y., Tsao, Y.: Generative adversarial networks for unpaired
  voice transformation on impaired speech. In: Interspeech. pp. 719--723 (2019)

\bibitem{chen2022phoneme}
Chen, X., Oshiro, A., Chen, J., Takashima, R., Takiguchi, T.: Phoneme-guided
  dysarthric speech conversion with non-parallel data by joint training.
  Signal, Image and Video Processing  \textbf{16}(6),  1641--1648 (2022)

\bibitem{chorowski2015attention}
Chorowski, J.K., Bahdanau, D., Serdyuk, D., Cho, K., Bengio, Y.:
  Attention-based models for speech recognition. Advances in neural information
  processing systems  \textbf{28} (2015)

\bibitem{chung2020defence}
Chung, J.S., Huh, J., Mun, S., Lee, M., Heo, H.S., Choe, S., Ham, C., Jung, S.,
  Lee, B.J., Han, I.: In defence of metric learning for speaker recognition.
  In: Interspeech. pp. 2977–--2981 (2020)

\bibitem{chung2018voxceleb2}
Chung, J.S., Nagrani, A., Zisserman, A.: Voxceleb2: Deep speaker recognition.
  arXiv preprint arXiv:1806.05622  (2018)

\bibitem{conneau2020unsupervised}
Conneau, A., Baevski, A., Collobert, R., Mohamed, A., Auli, M.: Unsupervised
  cross-lingual representation learning for speech recognition. arXiv preprint
  arXiv:2006.13979  (2020)

\bibitem{doshi2021extending}
Doshi, R., Chen, Y., Jiang, L., Zhang, X., Biadsy, F., Ramabhadran, B., Chu,
  F., Rosenberg, A., Moreno, P.J.: Extending parrotron: An end-to-end, speech
  conversion and speech recognition model for atypical speech. In: ICASSP
  2021-2021 IEEE International Conference on Acoustics, Speech and Signal
  Processing (ICASSP). pp. 6988--6992. IEEE (2021)

\bibitem{graves2006connectionist}
Graves, A., Fern{\'a}ndez, S., Gomez, F., Schmidhuber, J.: Connectionist
  temporal classification: labelling unsegmented sequence data with recurrent
  neural networks. In: Proceedings of the 23rd international conference on
  Machine learning. pp. 369--376 (2006)

\bibitem{hernandez2022cross}
Hernandez, A., P{\'e}rez-Toro, P.A., N{\"o}th, E., Orozco-Arroyave, J.R.,
  Maier, A., Yang, S.H.: Cross-lingual self supervised speech representations
  for improved dysarthric speech recognition. In: Interspeech. pp. 51--55
  (2022)

\bibitem{hori2017advances}
Hori, T., Watanabe, S., Zhang, Y., Chan, W.: Advances in joint ctc-attention
  based end-to-end speech recognition with a deep cnn encoder and rnn-lm. In:
  Interspeech. pp. 949–--953 (2017)

\bibitem{huang2021preliminary}
Huang, W.C., Kobayashi, K., Peng, Y.H., Liu, C.F., Tsao, Y., Wang, H.M., Toda,
  T.: A preliminary study of a two-stage paradigm for preserving speaker
  identity in dysarthric voice conversion. In: Interspeech. pp. 1329--1333
  (2021)

\bibitem{jia2018transfer}
Jia, Y., Zhang, Y., Weiss, R., Wang, Q., Shen, J., Ren, F., Nguyen, P., Pang,
  R., Lopez~Moreno, I., Wu, Y., et~al.: Transfer learning from speaker
  verification to multispeaker text-to-speech synthesis. Advances in neural
  information processing systems  \textbf{31} (2018)

\bibitem{kalchbrenner2018efficient}
Kalchbrenner, N., Elsen, E., Simonyan, K., Noury, S., Casagrande, N., Lockhart,
  E., Stimberg, F., Oord, A., Dieleman, S., Kavukcuoglu, K.: Efficient neural
  audio synthesis. In: International Conference on Machine Learning. pp.
  2410--2419. PMLR (2018)

\bibitem{kim2008dysarthric}
Kim, H., Hasegawa-Johnson, M., Perlman, A., Gunderson, J., Huang, T.S., Watkin,
  K., Frame, S.: Dysarthric speech database for universal access research. In:
  Ninth Annual Conference of the International Speech Communication Association
  (2008)

\bibitem{li2019jasper}
Li, J., Lavrukhin, V., Ginsburg, B., Leary, R., Kuchaiev, O., Cohen, J.M.,
  Nguyen, H., Gadde, R.T.: Jasper: An end-to-end convolutional neural acoustic
  model. arXiv preprint arXiv:1904.03288  (2019)

\bibitem{liutwo2024}
Liu, D., Lin, Y., Bu, H., Li, M.: Two-stage and self-supervised voice
  conversion for zero-shot dysarthric speech reconstruction. In: International
  Conference on Asian Language Processing (IALP). pp. 428--432. IEEE (2024)

\bibitem{nagrani2017voxceleb}
Nagrani, A., Chung, J.S., Zisserman, A.: Voxceleb: a large-scale speaker
  identification dataset. arXiv preprint arXiv:1706.08612  (2017)

\bibitem{panayotov2015librispeech}
Panayotov, V., Chen, G., Povey, D., Khudanpur, S.: Librispeech: an asr corpus
  based on public domain audio books. In: 2015 IEEE international conference on
  acoustics, speech and signal processing (ICASSP). pp. 5206--5210. IEEE (2015)

\bibitem{rudzicz2011production}
Rudzicz, F.: Production knowledge in the recognition of dysarthric speech.
  Ph.D. thesis (2011)

\bibitem{veaux2016superseded}
Veaux, C., Yamagishi, J., MacDonald, K., et~al.: Superseded-cstr vctk corpus:
  English multi-speaker corpus for cstr voice cloning toolkit  (2016)

\bibitem{wan2018generalized}
Wan, L., Wang, Q., Papir, A., Moreno, I.L.: Generalized end-to-end loss for
  speaker verification. In: 2018 IEEE International Conference on Acoustics,
  Speech and Signal Processing (ICASSP). pp. 4879--4883. IEEE (2018)

\bibitem{wang2020learning}
Wang, D., Liu, S., Sun, L., Wu, X., Liu, X., Meng, H.: Learning explicit
  prosody models and deep speaker embeddings for atypical voice conversion. In:
  Interspeech. pp. 4813--4817 (2021)

\bibitem{wang2022speaker}
Wang, D., Liu, S., Wu, X., Lu, H., Sun, L., Liu, X., Meng, H.: Speaker identity
  preservation in dysarthric speech reconstruction by adversarial speaker
  adaptation. In: ICASSP 2022-2022 IEEE International Conference on Acoustics,
  Speech and Signal Processing (ICASSP). pp. 6677--6681. IEEE (2022)

\bibitem{wang2020end}
Wang, D., Yu, J., Wu, X., Liu, S., Sun, L., Liu, X., Meng, H.: End-to-end voice
  conversion via cross-modal knowledge distillation for dysarthric speech
  reconstruction. In: ICASSP 2020-2020 IEEE International Conference on
  Acoustics, Speech and Signal Processing (ICASSP). pp. 7744--7748. IEEE (2020)

\bibitem{wang2024unit}
Wang, Y., Wu, X., Wang, D., Meng, L., Meng, H.: Unit-dsr: Dysarthric speech
  reconstruction system using speech unit normalization. In: ICASSP 2024-2024
  IEEE International Conference on Acoustics, Speech and Signal Processing
  (ICASSP). pp. 12306--12310. IEEE (2024)

\bibitem{wang2017tacotron}
Wang, Y., Skerry-Ryan, R., Stanton, D., Wu, Y., Weiss, R.J., Jaitly, N., Yang,
  Z., Xiao, Y., Chen, Z., Bengio, S., et~al.: Tacotron: Towards end-to-end
  speech synthesis pp. 4006--4010 (2017)

\bibitem{yunusova2008articulatory}
Yunusova, Y., Weismer, G., Westbury, J.R., Lindstrom, M.J.: Articulatory
  movements during vowels in speakers with dysarthria and healthy controls
  (2008)

\end{thebibliography}

\end{document}